\documentclass[aps,prd,showpacs,amsfonts]{revtex4}

\def\*{\star}
\begin{document}
\title {Negative Probability and Uncertainty Relations } 
\preprint{hep-th/0105226}
\author{ Thomas Curtright} 
 \email{curtright@physics.miami.edu}
\affiliation{Department of Physics, University of Miami, Box 248046,
          Coral Gables, Florida 33124, USA} 
\author{ Cosmas Zachos}
 \email{zachos@hep.anl.gov} 
\affiliation{High Energy Physics Division, Argonne National Laboratory,
          Argonne, Illinois 60439-4815, USA}
\date{May 23, 2001}
\begin{abstract}
A concise derivation of all uncertainty relations is given entirely
within the context of phase-space  quantization, without recourse to operator
methods, to the direct use of Weyl's correspondence, or to marginal 
distributions of $x$ and $p$. 
\end{abstract}
\maketitle

\title{Manuscript Title:\\with Forced Linebreak}

\author{Ann  Author}
 \altaffiliation[Also at ]{Physics Department, XYZ University.}
\author{Second Author}%
\affiliation{%
Authors' institution and/or address\\
This line break forced with \textbackslash\textbackslash
}%

Phase-space quantization is the third autonomous and logically complete
formulation of quantum mechanics beyond the conventional ones based on
operators in Hilbert space or path integrals  \cite{moyal,bayen,reviews}. It
is free of operators and wavefunctions: observables and matrix elements are
computed through phase-space integrals of c-number functions (``classical
kernels") weighted by a Wigner function (WF) \cite{wigner,reviews}. This is a
phase-space distribution function which is {\em not} positive semi-definite,
and constitutes the Weyl correspondent \cite{weyl} of the density matrix in
the conventional formulation,
\begin{equation}
f_{mn}(x,p) = 
\frac{1}{2\pi}\!\int dy e^{-iyp} \psi^*_m(x-\frac{\hbar}{2}y)
\psi_n(x+\frac{\hbar}{2}y)=f_{nm}^* (x,p).   
\end{equation}
Operators of the conventional formulation, when properly ordered (eg,
Weyl-ordered), correspond uniquely to phase-space classical kernel functions,
while operator products correspond to $\*$-products \cite{groen} of these
classical kernels, the $\*$-product being a noncommutative and associative
operation encoding quantum mechanical action. The above wavefunctions,
however, may be forfeited, since the WFs are determined, in principle, as the
solutions of the celebrated $\*$-genvalue functional equations
\cite{dbf,cfz,hug,cuz}. Connections to the original, operator, formulation of
quantum mechanics may thus be ignored.

Recent M-theory advances linked to noncommutative geometry and matrix models
\cite{natied} apply spacetime uncertainty  principles \cite{noncom} reliant on
phase-space quantization and the $\*$-product.  Transverse spatial dimensions
act formally as momenta, and, analogously to quantum mechanics, their
uncertainty is increased or decreased inversely to the uncertainty of a given
direction.

For classical (non-negative) probability distributions, expectation values of
non-negative functions are likewise non-negative, and thus result in standard
constraint inequalities for the constituent pieces of such functions. On 
the other hand, in phase-space quantization, the distribution functions are
 {\em non-positive-definite}, such as, in general, the {\em
quasi}-probability WF: it was interpreted early on by Bartlett
\cite{bartlett}, and later by Feynman \cite{feynman}, as a ``negative
probability function", with the proper non-negative marginal probabilities
upon projection to either $x$ or $p$ space. Hence, a frequent first question 
in phase-space quantization is how Heisenberg's standard quantum mechanical 
uncertainty relation arises for moments of such distributions.

To be sure, Moyal derived these
uncertainty relations, in his original formulation of quantum mechanics in
phase space, by careful analysis of conditioned and marginal probabilities.
Nevertheless, plain evaluations of expectation values of the c-number
variables  $\langle x^2 \rangle$, $\langle p^2 \rangle$, etc, do not evince 
constraints;  and the student of deformation quantization is left
wondering how $\hbar$ enters the constraint of such expectation values of
(c-number) observables when  the variables $x,p$ do not contain $\hbar$. How
do their moments manage to constrain each other by extracting $\hbar$ out of
the Wigner function?

The answer lies in 
Groenewold's associative $\*$-product \cite{groen},  
\begin{equation} 
\star \equiv ~ \hbox{{\Large 
$e^{  i   \hbar(\stackrel{\leftarrow }{\partial }_{x}
\stackrel{\rightarrow }{\partial }_{p}-\stackrel{\leftarrow }{\partial }_{p}
\stackrel{\rightarrow }{\partial }_{x})/2} $}}~, 
\end{equation} 
which is the cornerstone of phase-space quantization. 
 Its mechanics is reviewed in \cite{cuz,dbf,hansen}. 
An alternate, integral, representation of this product is \cite{neumann}
\begin{equation} 
f\star g=(\hbar \pi)^{-2} \int 
du dv dw dz ~f(x+u,p+v)~g(x+w,p+z)~  \exp \left({2i\over \hbar} 
\left( uz-vw \right) \right) ,
\end{equation}
which readily displays associativity. The phase-space trace is 
directly seen in this representation to obey 
\begin{equation}
\int\! dp dx ~ f\star g = \int\! dp dx ~ fg
=\int\! dp dx ~ g\star f ~ \label{Ndjambi}.
\end{equation}
The WF spectral properties \cite{moyal} are reviewed and
illustrated in \cite{cfz,cuz}. Eg, the trace-normalization condition, 
\begin{equation} 
\int dxdp  ~f_{mn}(x,p)=\delta_{mn}~ ,     \label{kurandye} 
\end{equation}
and the spectral orthogonality conditions \cite{dbf},
$ f_{mn}\* f_{kl}=\delta_{ml}f_{kn}/{2\pi\hbar}   $.
Given (\ref{Ndjambi}), it follows that  $\int 
dxdp ~f_{mn}(x,p)f^*_{lk}(x,p)= \delta_{ml}\delta_{nk} / {2\pi\hbar}$.
For complete sets of input wavefunctions, it also follows that 
\begin{equation}
\sum_{m,n} f_{mn}(x,p)f^*_{mn}(x',p')=\frac{1}{2\pi\hbar}
\delta(x-x')\delta(p-p')  ~.  \label{Loukobomo} 
\end{equation}
An arbitrary phase-space function $\varphi(x,p)$ can thus be expanded as
$\varphi(x,p)=\sum_{m,n} c_{mn}f_{mn}(x,p)$.

Here, a concise proof of all uncertainty relations is provided
completely within the autonomous framework of phase-space quantization, unlike
extant discussions of such correlation inequalities, which rely on the
operator formulation of quantum mechanics. It is stressed that, in the
following, {\em no} operators occur, only the $\*$-product operation, and $x$
and $p$ are c-numbers.
The controlling fact is that expectation values of arbitrary real $\*$-squares
 are positive semi-definite, even though the Wigner distribution $f(x,p)$ 
itself is not.  Specifically, for any complex phase-space function $g(x,p)$, 
and any (real) Wigner function $f(x,p)$ representing a pure state, 
the following inequality holds:
\begin{equation}
\langle g^{*}\star g\rangle = 
\int\! dp dx (g^{*}\star g) f   \geq 0  ~ \label{Otyikondo}.
\end{equation}
The $\*$ is absolutely crucial here, and its removal leads to violation of the
inequality, as can easily be arranged by choosing the support of $g$ to lie
mostly in those regions of phase-space where the Wigner function is negative.
(The only pure state WF which is non-negative is the Gaussian
\cite{hudson,reviews,cfz}). In Hilbert space operator formalism, this relation
(\ref{Otyikondo}) would correspond to the positivity of the norm.  By 
(\ref{Ndjambi}), $\int\! dp dx (g^{*}\star g) f =\int\! 
dp dx (g^{*}\star g)\star f $, ie inside a
phase-space integral an ordinary product can be extended to a $\star$-product,
provided it not be part of a longer string. Eg, the one $\star$-product of the
left hand side cannot be eliminated, because of the extra ordinary product
with $f$.

To prove the inequality (\ref{Otyikondo}), it suffices to recognize that, 
for a pure state, its (real) Wigner function can be expanded in 
a complete basis of Wigner $\star$-genfunctions of a convenient Hamiltonian,
\cite{cuz},  $f=\sum_{n,m} c^*_m c_n f_{mn} $, for complex 
coefficients $c_n$, s.t. $\sum_{n} |c_n|^2=1$ , to satisfy (\ref{kurandye}). 
Then, it follows that \cite{takabayasi,baker}  
\begin{equation}
f\star f= f/h  ~ \label{Nkayi}.
\end{equation}
Consequently, given the relations (\ref{Ndjambi}),
$(g\star f)^* = f\star g^*$,  and the associativity of the 
$\star$-product,
\begin{equation}
\int\! dp dx ~   (g^{*}\star g) f = 
h \int   dxdp\  (g^{*}\star g) (f\star f)  =
h\int   dxdp\ (f\star g^*)\star   (g\star f) = 
h\int   dxdp\ |g \star f |^2.
\end{equation}
This expression, then, involves a real non-negative integrand and is itself
positive semi-definite.
(Similarly, if  $f_1$ and $f_2$ are pure state WFs, the transition 
probability between the respective states \cite{reviews} is also 
manifestly non-negative by the same argument: $\int\! dp dx f_1 f_2 = 
({2\pi\hbar} )^2 \int   dxdp\  |f_1 \star f_2 |^2.$)

Given (\ref{Otyikondo}), correlations of observables follow conventionally from 
specific choices of $g(x,p)$. For example, to produce Heisenberg's 
uncertainty relation, one only need choose
\begin{equation}
g= a + bx+cp,
\end{equation}
for arbitrary complex coefficients $a,b,c$. The resulting positive 
semi-definite quadratic form is then
\begin{equation}
 a^*a  + b^* b \langle x \star x\rangle + c^* c \langle p\star p\rangle 
+(a^*b +b^* a) \langle x\rangle + (a^* c+c^* a) \langle p\rangle+ 
c^* b\langle p\star x \rangle+ b^* c \langle x\star p \rangle  \geq 0, 
\end{equation}
for any $a,b,c$.  The eigenvalues of the corresponding matrix are then 
non-negative, and thus so must be its determinant. Given 
\begin{equation}
x\star x=x^2, \qquad   p\star p=p^2, \qquad 
   p\star x= px  -i\hbar/2, \qquad  x\star p= px  +i\hbar/2, 
\end{equation}
and the usual 
\begin{equation}
(\Delta x )^2 \equiv  \langle (x-\langle x\rangle )^2 \rangle ,  
\qquad \qquad 
(\Delta p )^2 \equiv  \langle (p-\langle p\rangle )^2 \rangle ,  
\end{equation}
this condition on the $3\times 3$ matrix determinant amounts to
\begin{equation}
(\Delta x )^2 ~(\Delta p )^2 \geq \hbar^2/4 +\Bigl  (\langle 
(x-\langle x\rangle ) (p-\langle p\rangle) \rangle \Bigr  )^2, 
  \label{Impfondo}
\end{equation}
and hence 
\begin{equation}
\Delta x ~\Delta p \geq \hbar/2. 
\end{equation}

The inequality is saturated for a vanishing original integrand $g\star f=0$,
for suitable $a,b,c$, and when the last term of (\ref{Impfondo}) vanishes:
$x,p$ statistical independence, such as in a Gaussian ground state 
WF, $f_{00}=2h \exp (-(x^2+p^2)/\hbar) $.

More general choices of $g$ will likewise constrain as many observables as 
this function has terms ($-1$, if there is a constant term). For instance, 
for more general (real) observables $u(x,p),v(x,p)$, the resulting inequality 
is
\begin{equation}
\Delta u ~\Delta v \geq \frac{1}{2} 
\sqrt{|\langle u\star v-v\star u \rangle |^2 +
\langle ( u-\langle u\rangle) \star (v-\langle v\rangle)+ 
( v-\langle v\rangle) \star (u-\langle u\rangle) \rangle^2}.
\end{equation}
The minimum uncertainty is realized at 
$\langle u\star v  +v\star u   \rangle =2\langle u  \rangle \langle v  \rangle$,
with $g\star f=0$ for specific coefficients, ie,
\begin{equation}
\Bigl  (  \Delta u ~(v-\langle v\rangle) -ki
\Delta v ~(u-\langle u\rangle)                                       
\Bigr )  \star f=0,
\end{equation}
where $k$ is the sign of $i\langle u\star v - v\star u\rangle $. Solving 
such $\*$-equations is elaborated in \cite{cfz,hug,cuz,hansen}.

\phantom{.}

This work was supported in part by the US Department of Energy, 
Division of High Energy Physics, Contract W-31-109-ENG-38, and the NSF Award 
0073390.

\end{document}